\documentclass[
 amsmath,amssymb,
 preprint,
]{revtex4-1}
\usepackage{xcolor}
\usepackage{graphicx}
\usepackage{dcolumn}
\usepackage{bm}

\usepackage[utf8]{inputenc}
\usepackage{mathptmx}
\usepackage{subcaption}
\begin{document}

\title{
Learning from the Density to Correct Total Energy and Forces in First Principle Simulations}
\author{Sebastian Dick}
\email{sebastian.dick@stonybrook.edu}
\author{Marivi Fernandez-Serra}%
\affiliation{%
Physics and Astronomy Department, Stony Brook University, Stony Brook, New York 11794-3800, United States
}%
\affiliation{%
Institute for Advanced Computational Science, Stony Brook University, Stony Brook, New York 11794-3800, United States
}%

\begin{abstract}
We propose a new molecular simulation framework that combines the transferability, robustness and chemical flexibility of an ab initio method with the accuracy and efficiency of a machine learning model. The key to achieve this mix is to use a standard density functional theory (DFT) simulation as a pre-processor for the atomic and molecular information, obtaining a good quality electronic density. General, symmetry preserving, atom-centered electronic descriptors are then built from this density to train a neural network to correct the baseline DFT energies and forces. These electronic descriptors encode much more information than local atomic environments, allowing a simple neural network to reach the accuracy required for the problem of study at a negligible additional cost. The balance between accuracy and efficiency is determined by the baseline simulation. This is shown in results where high level quantum chemical accuracy is obtained for simulations of liquid water at standard DFT cost, or where high level DFT-accuracy is achieved in simulations with a low-level baseline DFT calculation, at a significantly reduced cost.

\end{abstract}
\maketitle

\section{Introduction}
The development of atomistic  simulation methods that make use of machine learning (ML) techniques to achieve 
close or beyond ab initio level accuracy
at a highly reduced computational cost is proving to be an efficient and robust alternative to traditional force field approaches \cite{behler2016perspective,nguyen2018comparison,zhang2018deep,chmiela2017machine,schutt2017schnet, hansen2015machine, smith2017ani}. This high level of activity is facilitated and supported by the availability and continuous generation of larger and improved data bases of highly accurate calculations or experiments both for molecular \cite{gomez2018automatic,ghahremanpour2018alexandria,ramakrishnan2014quantum} and solid systems \cite{kirklin2015open,winther2018catalysis,zakutayev2017high}. While in many cases, newly proposed methodologies and force-fields are only at the proof of concept stage, some of the most promising methods have already produced results that provide new physical insight into challenging problems \cite{hellstrom2019one}. Beyond predicting atomic structures of complex materials \cite{deringer2018realistic}, they have been used to address problems previously inaccessible to ab initio level calculations such as the structure of the surface of water, or
its temperature dependent dynamical properties using the MB-pol model \cite{Babin2013, Babin2014, Medders2014,PhysRevLett.121.246101,reddy2017temperature}.

 However, even if all these new force fields are facilitating
 the simulation of much larger systems and achieving longer dynamic
 scales with near density functional theory (DFT) accuracy \cite{Behler2007,zhang2018deep, zhang2018end, smith2017ani} or beyond \cite{chmiela2017machine,chmiela2018, schutt2017schnet}, this
 does not imply that DFT is becoming obsolete. There is still a need to obtain accurate energies, densities and  other physical observables that can be derived from first-principles methods such as DFT.
 Hence, developing approaches that aim to improve DFT by searching for more accurate exchange and correlation potentials using information from  correlated wave function methods is a worthwhile endeavour. It is known that DFT-based methods within their lower rungs of approximation to the exchange and correlation (XC) energy 
 struggle in simulating systems like water or ionic solutions \cite{distasio2014individual, gillan2016perspective,yao2018free,chen2017ab}. The use of more expensive XC approximations, such as hybrid functionals \cite{becke1993density}, may deliver the required accuracy, however the size of systems for which these simulations are feasible is limited.
 This trade-off between accuracy, flexibility and computational cost seems to lie at the heart of many problems the molecular science community is presently facing. 

Some progress has been made in the development of machine-learned density functionals. Apart from promising efficient orbital-free DFT calculations by learning the kinetic energy functional \cite{seino2018semi, snyder2012finding, yao2016kinetic}, these methods have been shown to allow to skip self-consistent calculations altogether by directly learning the map from atomic potential to electron density \cite{brockherde2017bypassing, grisafi2018transferable}. While these applications still appear to be limited to one-dimensional systems or small molecules, other
approaches have shown promising results in resolving this size extensivity problem. \cite{lei2019design,doi:10.1063/1.5029279,nagai2019completing,chandrasekaran2019solving}. In this work, we propose an alternative approach called Machine Learned Correcting Functionals (MLCFs) which, rather than replacing DFT, complements it by adding a machine learned functional of the electron density,  expanded into a set of localized basis functions. Even though not explicitly shown here, this machine-learned density functional, based on local electronic descriptors, can in principle be used in self-consistent calculations. Exploring this path will be the topic of future work. The principal concept of this new methodology could be termed ``Informed Machine Learning.'' We argue that all the available physical insight into a problem should be used to help the ML-model make its predictions and minimize its generalization error. In MLCFs, this translates to DFT assuming the role of a data-preprocessor that maps the atomic coordinates to a self-consistent electron density which is input into an artificial neural network (ANN). The ANN is then fitted to the difference in energy and forces between the baseline method which was used to obtain the electron density and a higher-level reference method of choice.
Thus MLCFs draw upon and optimize the strengths of DFT, which is effective in calculating mean field electrostatic interactions, but often fails to accurately describe non-local interactions rooted in quantum mechanics. At the same time, DFT provides a solid foundation which the ML-model can fall back on when faced with unseen scenarios in which it will likely fail to make reliable predictions.

All results presented in this work were obtained with our MLCF implementation, available at Ref. \cite{mlcf_code}.

\section{Methods}

\subsection{Charge density representation}
As in any machine learning application, the representation of the data used for input (aka, the features), should encode as much information in as few variables as possible, to avoid redundancies and make sure that the ANN model is quick to train and not prone to overfitting. Furthermore, we require the number of descriptors to scale linearly with the number of atoms in the system to guarantee size extensivity of the energy. 

To ensure that the model respects rotational symmetry we further impose the constraint that descriptors need to transform covariantly under $SO(3)$ rotations.

One way to create such a descriptor is to project the electron density in real space onto a set of atom-centered orbitals, which were in inspired by Bartok et al.'s work on representing chemical environments \cite{bartok2013representing}. While completing this manuscript, the authors became aware of the work by Grifasi et al. \cite{grisafi2018transferable} which uses a very similar approach.

The angular and radial basis functions are given by spherical harmonics $Y^m_l(\theta,\phi)$ and
\begin{align}
&\tilde\zeta_n(r) = \left\{
                \begin{array}{ll}
                \frac{1}{N}(r-r_i)^2(r_o -r)^{n+2}
&\text{ for } r > r_i \text{ and } r < r_o \\
0 &\text{ else }
		\end{array}
			      \right.
\end{align} 
respectively.
{ Here we have used a radius $r_o$ and a normalization factor $N$, the latter being determined by numerical integration. {The radial functions defined above correspond to the ones used by Bartok et al. \cite{bartok2013representing} except for the addition of an inner cutoff radius $r_i$ which we included to disregard the core area. Moreover, the exponents were chosen so that the basis has vanishing first and second derivative at both cutoff radii to ensure smoothness at these points. We found that this somewhat mitigates artifacts arising from the discrete nature of the euclidean grid used to represent the electron density.}}

\begin{table}[t]
 \begin{tabular}{c|c|c}
  & Oxygen & Hydrogen \\
  \hline
  \hline
  $n_{rad}$ & 2 & 2  \\
  $l_{max}$ & 2 & 1  \\
  $r_i$ & 0.05 & 0.0  \\
  $r_o$ & 1.0 & 1.5  \\
 \end{tabular}
 
\caption{Hyperparameters used to create the charge density descriptors from top to bottom: Number of radial functions, maximum angular momentum, inner radial cutoff, outer radial cutoff . Radii are given in units of Angstrom.}\label{tab:hyperparameters}
\end{table}

After orthogonalizing the radial functions with the transformation 
\begin{align}     
&\zeta_n(r) = \sum_{n'} S^{-1/2}_{nn'} \tilde \zeta_n(r), \quad \text{where } S_{nn'} = 
\int dr \tilde\zeta_n(r) \tilde\zeta_{n'}(r).
\end{align}

the full basis is given as $\psi_{nlm}(\vec{r}) = Y^m_l(\theta, \phi) \zeta_n(r)$.

The descriptors $c^{\alpha,I}_{nlm}$ for atom $I$ of species $\alpha$ at position $\vec r_{I}$ can be obtained by projecting the electron density $\rho$ onto the corresponding basis functions $\psi^\alpha_{nlm}$.

\begin{equation}
 \tilde c^{\alpha,I}_{nlm} = \int_{\vec r} \rho(\vec r - \vec r_{\alpha, I}) \psi^{*\alpha}_{nlm}(\vec{r}). \label{eq:descr}
\end{equation}
Introduction of an atomic species label $\alpha$ has made it possible to have different bases for distinct atomic species.

To ensure rotational invariance, the basis should be aligned with a uniquely defined local coordinate system (LCS).
Knowing the Euler angles $\{\alpha,\beta,\gamma\}$ that 
relate the global coordinate system (GCS) to the LCS, the rotated descriptors are given as
\begin{equation}
 c_{nlm} = \sum_{m'} (D^{l}(\alpha,\beta,\gamma))^{-1}_{mm'} \tilde c_{nlm'},
\end{equation}
with $D^l$ being the Wigner D-matrix for angular momentum {\cite{wigner1939unitary,thomas2018tensor,cohen2016group}}.
The LCS of a given atom can be defined trough the position of atoms in its immediate environment.

Even though this approach has been proven to work for machine learned force fields \cite{zhang2018deep},
it would be desirable to have a definition of the LCS that is completely independent of the atomic coordinates and only relies on electronic information.  

Recognizing that the descriptors associated with the p-orbitals $\tilde c_{n1m}$ transform like vectors under $SO(3)$ rotations, we can use these vectors to define a local coordinate system that only depends on the electron density (see appendix for more details). 

\begin{figure}[t]
 \includegraphics[width=.45\textwidth]{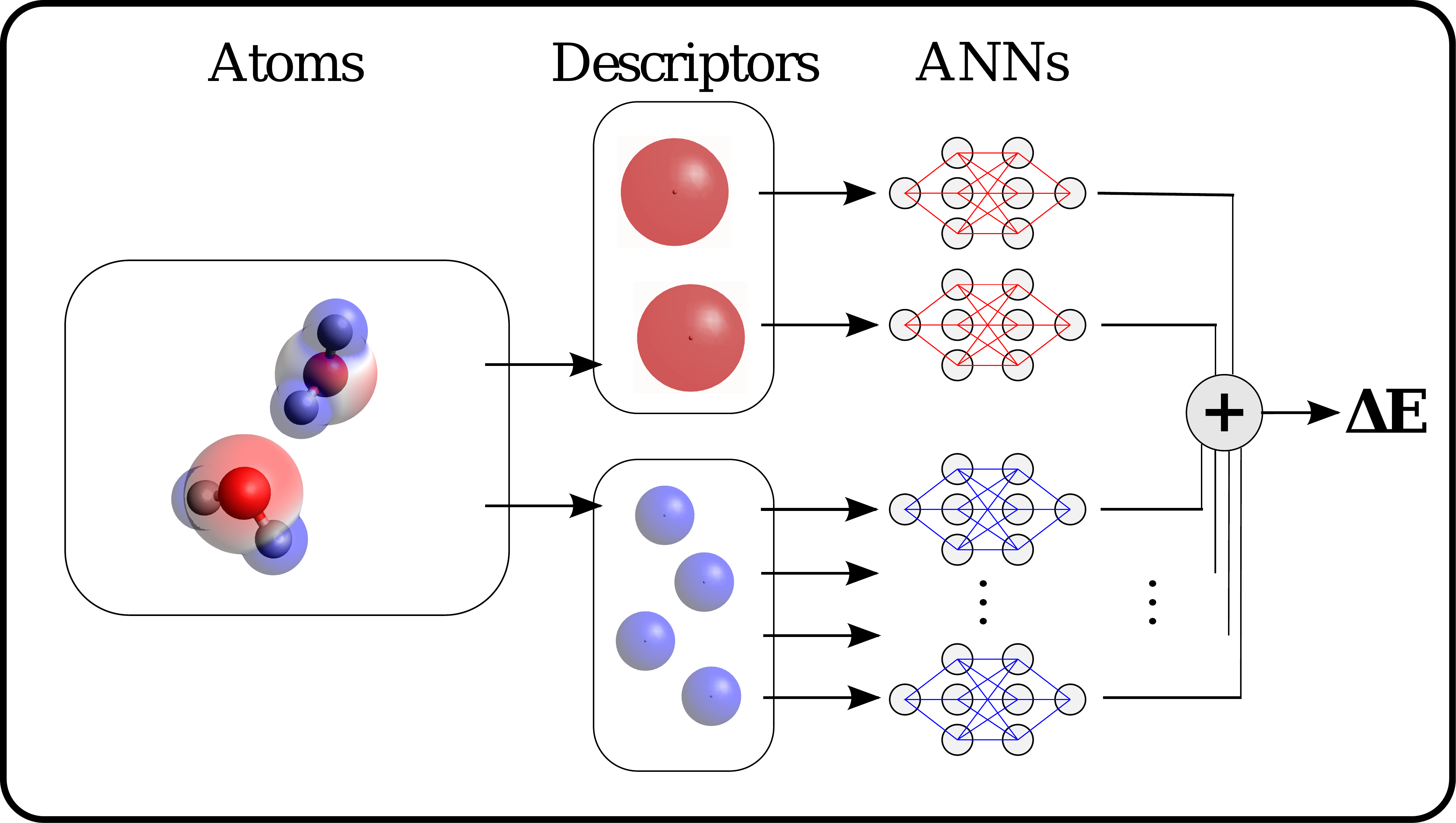}
 
  \caption{Architecture of a Behler-Parrinello type neural network. Starting with a given configuration, the self-consistent electron density is computed with the baseline DFT. The density is then encoded in descriptors that are fed into artificial neural networks (ANNs) which calculate atomic contributions to the total energy correction $\Delta E$.}\label{fig:model}
\end{figure}

Aligning the descriptors independently of structural variables has the major advantage that conformers that are close in their electron densities are described by similar features. This is particulary important for MLCFs that only correct the exchange correlation energy $E_{xc}$, as these conformers will necessarily exhibit the same error in $E_{xc}$.

The ideal number of radial and angular basis functions as well as the inner and outer cutoff radii can be determined by cross-validation. Doing so, the hyperparameters listed in Tab. \ref{tab:hyperparameters} were found to provide a good compromise between accuracy and computational efficiency.

\subsection{Machine learning models} 
Inspired by the well-established Behler-Parrinello networks \cite{Behler2007}, the energy functional for a given number of molecules 
was represented as an artificial neural network (ANN) consisting of a sum of smaller atomic networks, each of which     
only sees the local charge density around the atom it is associated with (see Fig.\ref{fig:model}). Furthermore,
networks belonging to atoms of the same species are identical to guarantee permutational invariance. Separate models were trained directly on the force corrections to be used in molecular dynamics simulations. Details about the force models are provided in the appendix .

Both models were trained using the Adam \cite{kingma2014adam} optimizer with training rate $\alpha = 0.001$ and decay rates $\beta_1 = 0.9$ and $\beta_2 = 0.999$. { The validation set was created by splitting off 20\%  from the original training set}, and the number and size of layers as well as the training rate was determined using cross-validation. { No mini-batches were used for training, i.e. the entire training set was passed through the neural network for every parameter update. The maximum number of epochs was set to 40000 but due to early stopping, this value was never reached. Typical numbers of training epochs ranged between 10000 and 30000 and training times were of the order of minutes}. A network with three layers of 8 nodes and one with three layers of 16 nodes were determined to have the optimal accuracy for energy and force predictions respectively, however satisfying results were obtained with a wide variety of model sizes.

The sigmoid function was chosen as activation 
\begin{equation}
 f(x) = \frac{1}{1+e^{-x}}
\end{equation}
as it was determined {through cross-validation} to be less prone to over-fitting than other possible choices such as the hyperbolic tangent or the rectified linear unit. { One possible problem that practitioners should be aware of when using sigmoid or related functions as activations is what is commonly known as the vanishing gradient problem. The vanishing gradient problem stems from the fact that for very large and small input values the derivative of the sigmoid function vanishes, which makes training exceedingly difficult for networks with many layers. This is usually mitigated by using rectified linear units (ReLU) as activation in deep neural networks \cite{goh2017deep}. We have  not encountered any problems during the training process, which is most likely due to the limited depth of our neural network.}

\begin{figure}[t]
 \includegraphics[width=.43\textwidth]{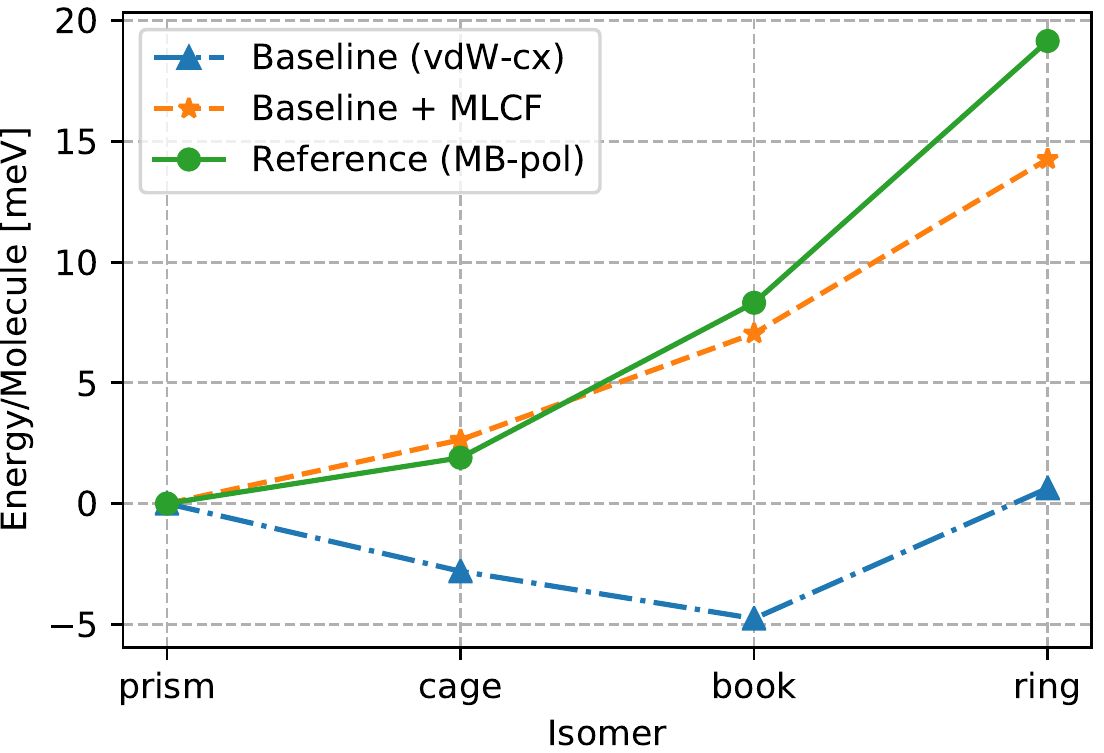}
  \caption{Energies of water hexamers with respect to the prism isomer, which is correctly determined to be the most stable structure by the MLCF.}\label{fig:hexamers}
\end{figure}

\section{Results}

\subsection{Gas phase water} 
We tested our method on water clusters of varying size. The baseline calculations were conducted using SIESTA \cite{soler2002siesta} with a quadruple zeta doubly polarized basis set and a van der Waals density functional of Dion et al. \cite{dion2004van} with GGA exchange modified by K. Berland and P. Hyldgaard (vdW-cx) \cite{berland2014exchange}. This XC functional has been shown to be as good as other vdW-type functionals in describing liquid water\cite{fritz2016optimization}.

For the reference energies, we use the MB-pol force field \cite{Babin2013, Babin2014, Medders2014} which is fitted to highly accurate coupled cluster calculations of water monomers and clusters. MB-pol has been shown to accurately reproduce the structural and thermodynamic properties of condensed phases of water, hence it is a superior model to any vdW or GGA XC functional. It is important to emphasize that our method is of course not limited to the use of force-fields as a reference method. As only moderately sized data sets of relatively small systems are needed, high level quantum chemistry calculations to produce these data sets are expensive but feasible. 

 The MLCF was trained to interpolate between the baseline and reference method, i.e. the target values used for fitting are given as the differences in energy $\Delta E = E_{ref} - E_{base}$.

The data set comprised about 400 water monomers, 2000 dimers and 1500 trimer configurations that were all sampled from the data that was used to fit MB-pol \cite{Babin2013, Babin2014, Medders2014}, and an additional 300 monomers with a uniform distribution of bond lengths and angles. 

The data was split into 80 percent training and 20 percent hold-out set.
We further created samples of larger water clusters with $n = 4,5,8,16$  molecules (50 samples each) for testing the size extensivity of the network. 

In Tab. \ref{tab:cluster_energies} the errors in relative energies of isomers are split up into root mean square (RMS) error, mean absolute error (MAE) and maximum (absolute) error.

\begin{table}[h]

\begin{tabular}{c|c|c|c}
\toprule
No. Molecules &          RMSE &           MAE &      max. Error \\
\hline
1         &  2.29 (53.19) &  1.25 (43.13) &  14.71 (151.19) \\
2         &  4.33 (40.17) &  2.91 (31.28) &  31.03 (136.91) \\
3         &  2.89 (28.25) &  2.16 (22.29) &   12.27 (75.20) \\
4         &   2.79 (9.69) &   2.15 (7.93) &    7.70 (24.95) \\
5         &  2.64 (11.24) &   2.19 (8.97) &    6.23 (36.69) \\
8         &   3.43 (9.26) &   2.72 (7.34) &    8.05 (22.67) \\
16        &   2.75 (6.28) &   2.15 (5.03) &    6.19 (17.15) \\
\hline
\end{tabular}
\caption{Model performance on test sets containing clusters of different sizes. Reported are the root mean square error (RMSE), mean absolute error (MAE) and, maximum absolute error (max. Error), given in meV/Molecule. The baseline method errors are given in parenthesis. } \label{tab:cluster_energies}
\end{table}
\begin{figure}[t]
 \includegraphics[width=.47\textwidth]{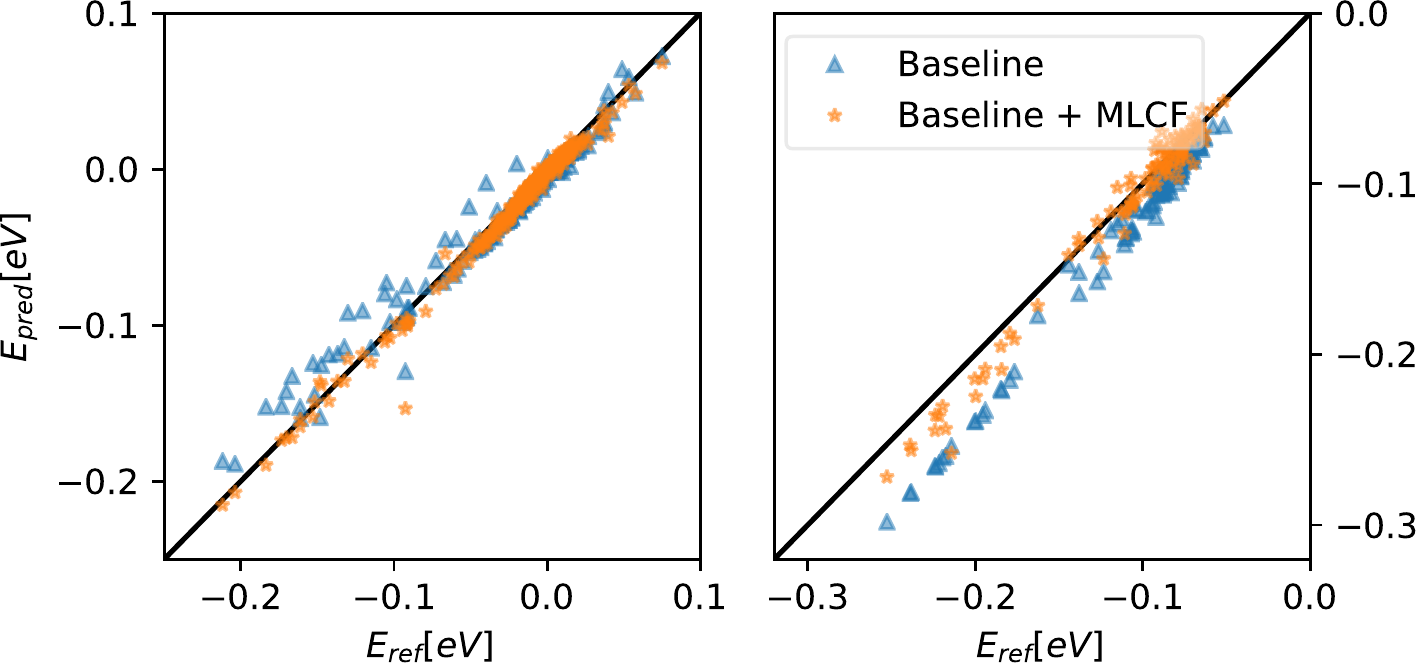}
  \caption{Predicted vs. reference two (left) and three (right) body energies for the data sets taken from ref.  \citealp{fritz2016optimization}. For the two body energies the RMSE is 4.20 for the baseline method and 2.80 for the MLCF corrected method. For the three body energies the errors are 7.10 and 3.56 respectively. Errors are given in meV/Molecule.}\label{fig:nbody}
\end{figure}

As can be seen in Tab. \ref{tab:cluster_energies}, in spite of only being trained on systems with up to three molecules, our model performs well for all cluster sizes and significantly 
decreases errors in relative energies of isomers, all RMS errors being below 5 meV/Molecule and mean absolute errors below 3 meV/Molecule.
Moreover, Fig. \ref{fig:hexamers} shows that the overall energy ordering of hexamer structures is correctly reproduced by the MLCF, a task that has so far proven largely elusive to GGA-level DFT calculations \cite{fritz2016optimization}.

To gain further insights, we can follow the approach of previous works \cite{bartok2013machine, fritz2016optimization} and split up the total energy of water clusters into their n-body contributions. 
This was done on the same data sets that were used in ref. \citealp{fritz2016optimization} to optimize the GGA exchange enhancement factor for water systems.

Figure \ref{fig:nbody}, shows that our model not only corrects the predominant errors in the one body energy (Tab. \ref{tab:cluster_energies}) but also improves two and three body energies. 
This might seem surprising as the MLCF is only provided with the local charge density around each atom. However, the model seems to learn how to infer higher order interaction energies from correlations in the local charge density. This indicates that long range, and possibly non local interactions are encoded within local density spatial correlations, enabling us to use radial cutoffs for the descriptors that are very short {(see Tab.  \ref{tab:hyperparameters})}, comparable to the nearest neighbor distance, { which is about 1 \r{A} in water molecules}. In order to obtain similar accuracy using only structural information, much larger cutoffs are needed \cite{bartok2013representing,bartok2013machine}.

{
\subsection{Comparison with machine learned force fields}

Rather than competing with state of the art machine learned force fields our work is aimed at investigating possible paths towards machine learned functionals that use the eletron density as input. Nevertheless, it is instructive to compare the performance of our model to ML methods that make predictions using only geometric information. In particular we want to investigate their capability to extrapolate, which restricts our possible choices to methods that learn a size extensive representation of the total energy. This disqualifies models that assume a fixed input size such as the sGDML approach by Chmiela et al. \cite{Chmiela2016,Chmiela2017,chmiela2018}. Out of the plethora of publicly available ML force-fields we chose to compare our method to SchNet \cite{schutt2017schnet, schutt2018schnet} and a Behler-Parrinello network (BPN)\cite{Behler2007} with weighted atom-centered symmetry functions (WACSF) \cite{gastegger2018wacsf} both implemented in the open-source library SchNetPack \cite{schutt2018schnetpack}.

For a fair comparison, we require these force fields to merely learn the difference between the reference (MB-pol) and baseline (vdW-cx) energies and we will therefore refer to the them as $\Delta$-SchNet and $\Delta$-WACSF from here on. In analogy to our analysis for MLCF we want to test how a model trained on monomers, dimers and trimers extrapolates to larger water clusters. The performance of non-$\Delta$ models, i.e. models that are trained on total energies is shown in  Fig. \ref{fig:total} in the appendix . All non-$\Delta$ models trained show generalization errors above 10 meV/$H_2O$ and can therefore not compete in accuracy with MLCF or their $\Delta$-learning counterparts.

It is not clear, a priori, which is the appropriate cutoff distance for the force fields. 
For non-$\Delta$ models it is obvious that the cutoff should be chosen equal to the largest atomic distance in the training set, which is about 8 \r{A}. Intuitively we would expect that for $\Delta$-models a smaller cutoff will be sufficient and maybe even beneficial for the model's capability to extrapolate. Instead of fixing the radial cutoff we opted to train a collection of models with cutoff ranging from 2 to 8 \r{A}. For every choice of cutoff, the optimal set of hyperparameters was determined by a grid-search and cross-validation (see appendix  for details).   

\begin{figure}[t]
 \includegraphics[width=0.48\textwidth]{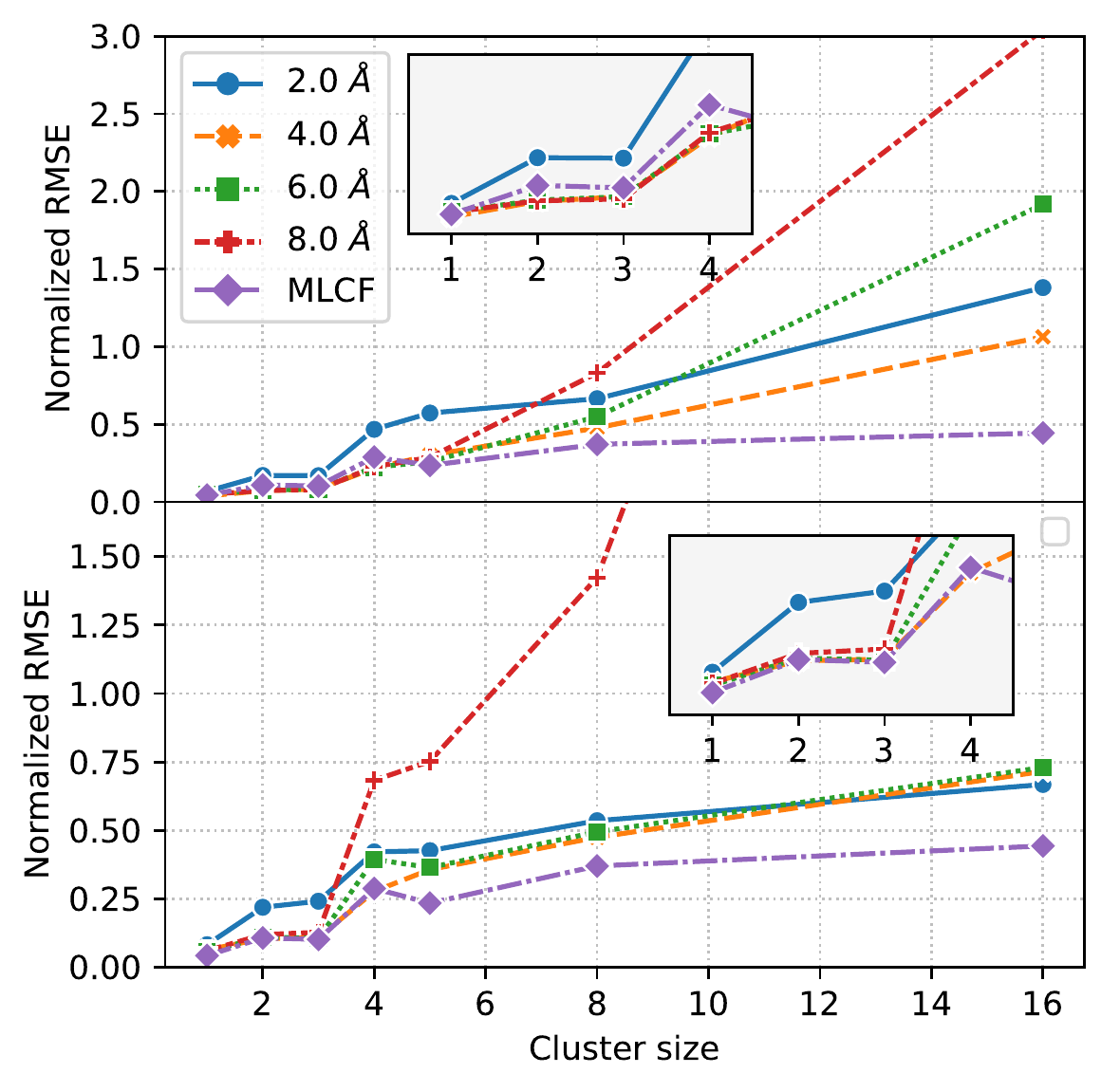}
  \caption{Root mean squared errors (RMSE) of $\Delta$-SchNet (top) and $\Delta$-WACSF (bottom) models trained for different cutoff radii (see legend) on water monomers, dimers and trimers and tested on random clusters of increasing size. The errors are normalized by dividing them through the RMSE of the underlying baseline method (vdW-cx) and compared to those of MLCF (purple line) which can also be found in Tab. \ref{tab:cluster_energies}. }\label{fig:diff}
  
\end{figure}

As can be seen in Fig. \ref{fig:diff}, for both $\Delta$-SchNet and $\Delta$-WACSF a cutoff of 4 \r{A} results in the best extrapolation to larger clusters.  $\Delta$-SchNet  outperforms $\Delta$-WACSF for small clusters but performs worse in terms of extrapolation. $\Delta$-SchNet is slightly more accurate than MLCF for cluster sizes below 5 molecules but both force-fields are less capable of extrapolating to large clusters than MLCF. It is instructive to see that using the electron density as input, smaller cutoff radii (1.0 \r{A} - 1.5 \r{A}) are sufficient to reach competitive accuracy. This suggests that the network learns to infer how non-local information is encoded in the local charge density. 

To go one step further, we would like to compare the models' performance in systems with other molecules present. For this, we used the publicly available data set S66x8, which contains dissociation curves for 66 non-covalent complexes relevant to biomolecular structures \cite{rezac2011s66}. We limited our calculations to a subset with complexes that contain at least one water molecule. For $\Delta$-SchNet and $\Delta$-WACSF we used the models with cutoff 4 \r{A} from above, the MLCF remained unchanged to the one used for water clusters.

By design, the MLCF is only ever used for atomic species that it was trained on, i.e. Oxygen and Hydrogen. For the force-fields, best results were achieved if corrections were restricted to these elements as well. Therefore, when using these force-fields, we masked out all other atoms before creating the input features. A comparison of all three methods for a selected subset can be seen in Fig. \ref{fig:s66}. None of the methods correctly predicts the dissociation curve for water and methylamine (Fig. \ref{fig:s66}d). This is due to the fact that the H-N hydrogen bond was not contained in the training set. In fact the nitrogen is 'invisible' to the machine learning method in all cases. 
For the systems containing methanol, $\Delta$-WACSF and MLCF both make reasonable predictions, the latter being more accurate than the former in the case where methanol acts as the donor molecule. In all cases $\Delta$-SchNet shows deviations for large distances and around the equilibrium distance.

The most instructive example is given by dissociation curve for  n-methylacetamine and  water. The crucial difference to the systems above is that now the Hydrogen participating in the bond is covalently bonded to a Nitrogen atom. In terms of the electron density the changes to OH are small, which explains why MLCF still performs well. However, making use of only structural information, both force-fields perform worse than the baseline method in this case.

\begin{figure}[t]
 \includegraphics[width=0.48\textwidth]{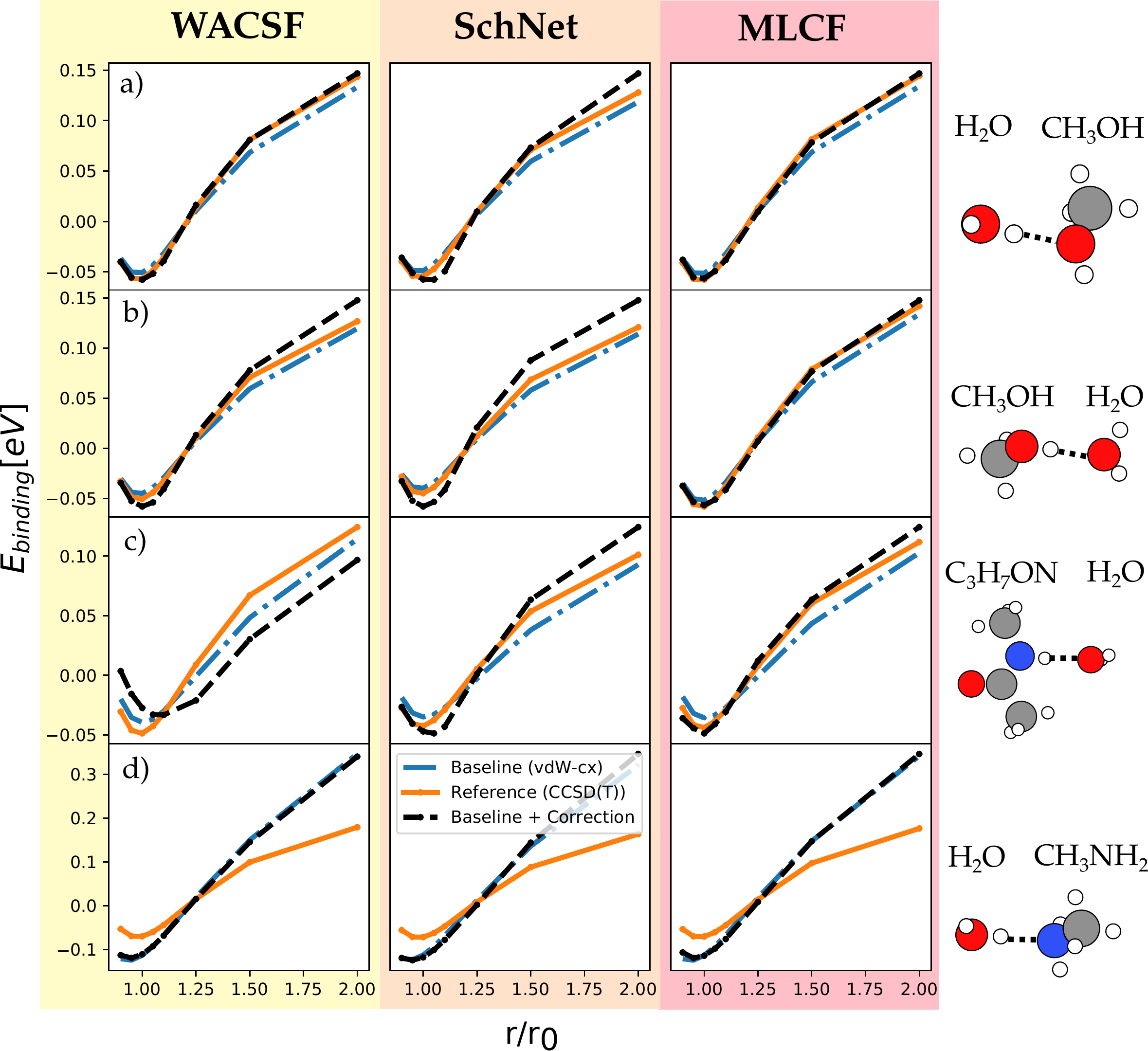}
  \caption{Comparison of, from left to right, $\Delta$-WACSF, $\Delta$-SchNet and MLCF on a subset of dissociation curves taken from the s66x8 \cite{rezac2011s66} data set. From top to bottom the figures describe hydrogens bonds between: a) water and methanol, b) methanol and water, c) n-methylacetamine ($C_3H_7ON$) and water, and  d) water and methylamine ($CH_3NH_2$), where the molecule listed first is understood to be the hydrogen donor. The energies depicted correspond to binding energies for the reference calculations (orange) and total energies for the baseline (blue) and corrected (black) calculations. As the molecules stay rigid during the relative displacement, binding and total energies merely differ by a constant. We therefore shifted every curve by its mean energy to align them and draw better comparisons between the methods. The depicted distances were normalized by dividing them through their respective equilibrium values, which were determined with MP2.}\label{fig:s66}
\end{figure}
}

\subsection{Liquid water} 
Having proven successful for gas-phase calculations, we went on to assess the method's performance in molecular dynamics simulations on liquid water. 
For reasons of practicality we opted to replace the previously employed van der Waals functional with the faster Perdew-Burke-Ernzerhof (PBE) \cite{perdew1996generalized} functional.
We trained a MLCF to interpolate between this baseline method and MB-pol, which has been shown to produce pair correlation functions close to experimental results. Further details about the training data set used are given in the appendix .

\begin{figure}[t]
 \includegraphics[width=.45\textwidth]{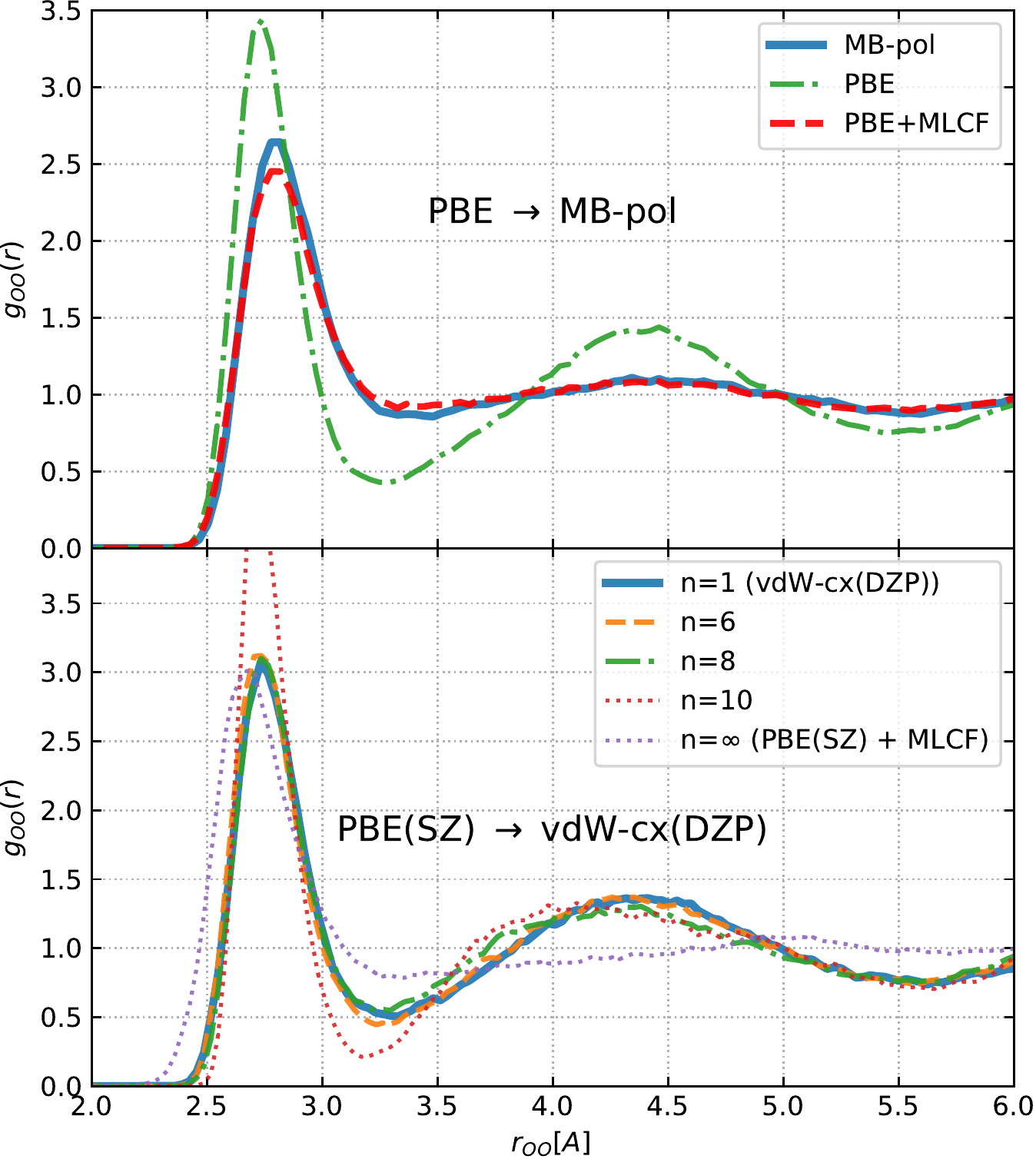}
  \caption{Oxygen-oxygen radial distribution functions $g_{OO}(r)$ for two different applications of MLCFs. \textit{Top}: Correcting for errors in the XC functional by interpolating between PBE and MB-pol. \textit{Bottom}: Accelerating DFT calculations by correcting for both functional and basis-set related convergence errors, combining MLCF with the time step mixing method proposed by Anglada et al. \cite{anglada2003efficient}. Interpolation between a quick and dirty PBE(SZ) simulation and a slow and more accurate vdW-cx(DZP) simulations. $n$ indicates the number of fast steps needed before correcting with a well converged step.}\label{fig:md}

\end{figure}

We conducted a DFT-based molecular dynamics (MD) simulations of 128 water molecules at experimental density in a periodic box using a Nose-Hoover thermostat  \cite{martyna1992nose}  at a temperature of 300 K. The simulation length was 10 ps with a timestep of 0.5 fs using 4 ps for equilibration and 6 ps to create the correlation functions. As can be seen in the top section of Fig. \ref{fig:md}, our MLCF provides a significant improvement over PBE, which is known to over-structure liquid water.
The MCLF approach also corrects the dynamical properties of the baseline calculation, bringing them to the level of the reference method. This can be seen in the vibrational density of states plots shown in the appendix .

\subsection{Basis set correction} 

Beyond correcting the density functional approximation, MLCFs can also be used to extrapolate between two given basis sets. Thus, extremely fast calculations can be performed with minimal (single zeta) basis-sets that with the help of MLCFs are then brought to the accuracy of more conservative basis sizes.

Here it should be noted that MLCFs rely upon the quality of the electron density used as input.
Following an approach by Kim et al. \cite{kim2013understanding}, the error that the MLCF is trained to correct can be decomposed as  
\begin{equation}
    \Delta E = E_{ref}[n_{ref}] - E_{base}[n_{base}] = \Delta E_F  + \Delta E_D
\end{equation}
with $E_F = E_{ref}[n_{base}] - E_{base}[n_{base}]$ and $E_D = E_{ref}[n_{ref}] - E_{ref}[n_{base}]$ being the functional and density driven error respectively.
By design, MLCFs are only effective at correcting functional driven errors { as deviations in the density induce long-range effects through electrostatic interactions that are hard to correct with our semi-local approach}. This generally does not pose any serious problems, as most density functional approximations yield a good quality electron density. Using very sparse basis sets, however, imposes an upper limit on the accuracy of the density, making density driven errors become dominant.

{ It turns out that in spite of the inaccuracies produced by the small basis set, we can still obtain satisfactory results in MD simulations by using a method developed by Anglada et al.} \cite{anglada2003efficient}. In their paper, they propose to alternate fast steps calculated with a cheaper, less accurate method with slow steps obtained with a well converged method. The forces used to integrate the equations of motion with a time step $\Delta t$ are given as $\vec F(t) = \vec F_{fast}(t) + \Delta \vec F(t)$ where the correction term $\Delta \vec F(t)$ is defined as 
\begin{equation}
\Delta \vec F = 
\left\{\begin{array}{ll}
   n \delta \vec F  & \text{if } (t/\Delta t\mod n = 0)  \\
   0  & \text{else}
\end{array}\right .
\end{equation} with $\delta \vec F =  \vec F_{slow} - \vec F_{fast}$.
The motivation behind this idea is that the error $\delta \vec F$ is small compared to the absolute forces and relatively smooth over time. Therefore, it is sufficient to only correct for this error every $n$ steps, the limiting value of $n$ being dependent on both the system and the accuracy of the fast method.

To test this, we ran AIMD simulations of 64 water molecules, at 300K and experimental density for 10 ps.

The fast steps were calculated using the PBE exchange-correlation functional with a single zeta basis set and subsequently corrected with an adequately trained MLCF. For the slow, well-converged steps we used the vdW-cx functional and a double zeta polarized basis set.
{ The MLCF was trained to lift the accuracy of the fast method to that of the slow one. For that, we used the same procedure and data set as for the gas-phase MLCF discussed above. However we re-did all the DFT calculations replacing vdW-cx by PBE with single zeta basis set and MB-pol by vdW-cx with a polarized double zeta basis set, adjusting baseline and reference methods to fit the problem at hand.}  

As can be seen in the bottom part of Fig. \ref{fig:md}, the radial distribution function remains unchanged for $n<=6$ and reasonable results can be achieved up to $n<=8$.

It is important to point out that these results could only be obtained by correcting the fast steps with an MLCF. In its original form, even though proven successful for systems like liquid silica \cite{anglada2003efficient}, the mixing method failed for water. 
Furthermore, the upper limit for the number of subsequent fast steps $n_{lim}$ is strongly system-dependent and rigorous testing should be conducted before using this mixing scheme.
However, if carefully used, this method promises significant speed-ups that are roughly equal to $n_{lim}$ for large system sizes (see appendix  for detailed analysis).

\section{Conclusion}
In conclusion, we have presented and tested a new molecular simulation framework in which the electronic density is used to train a neural network to correct baseline DFT energies and forces to the accuracy provided by a higher level method.
Our results indicate that real space semilocal correlations of the electronic density as encoded in our localized density desriptors,
contain information about longer range effects. These effects can be corrected through the use of a neural network trained on purely local data, an observation that might support the
potential for developing machine learned density functionals using local density descriptors \cite{grisafi2018transferable} and total energy targets.

{
The scope of this method is not the same as that of machine learned force fields, given that it aims to stay within the realm of DFT. However, we have been able to show that indeed the use of local descriptors built from the electronic density outperforms those methods in terms of transferability. This suggest that electronic descriptors such as the ones proposed here could be explored to improve the accuracy of force fields.

After completion of this manuscript, the authors became aware of work by Bogojeski et al. \cite{bogojeski2019density} that is very closely related to the work presented here. Similar to our approach, Bogojeski et al. use density-based inputs and $\Delta$-learning to achieve quantum chemical accuracy. The main difference between the two approaches lies in the choice of basis functions and the way symmetries are encoded. In Bogojeski et al.'s method, the molecule is first aligned with a global coordinate system which is defined through some molecular axes. The electron density is subsequently expanded in a Fourier basis. These design choices seem to restrict their method to systems of fixed size and results presented are limited to small and medium sized molecules. In contrast, as we have shown above, by using semi-local, atom-centered basis functions, our method is size-extensive and can therefore extrapolate to systems it was not explicitly trained on, in particular condensed phases.}

\begin{acknowledgments}
We aknowledge funding from DOE awards numbers DE-FG02-09ER16052 and DE-SC0019394.
{
Sebastian Dick was supported by a fellowship from
The Molecular Sciences Software Institute under
NSF grant ACI-1547580.
We would like to thank Stony Brook Research Computing and Cyberinfrastructure, and the Institute for Advanced Computational Science at Stony Brook University for access to the high-performance SeaWulf computing system, which was made possible by a \$1.4M National Science Foundation grant (\#1531492).
Finally, we want to express our thanks to Jose Soler for valuable advice concerning this project.
}
\end{acknowledgments}

\appendix
\section{Coordinate systems}
\begin{figure}][t]
 \includegraphics[width=.4\textwidth]{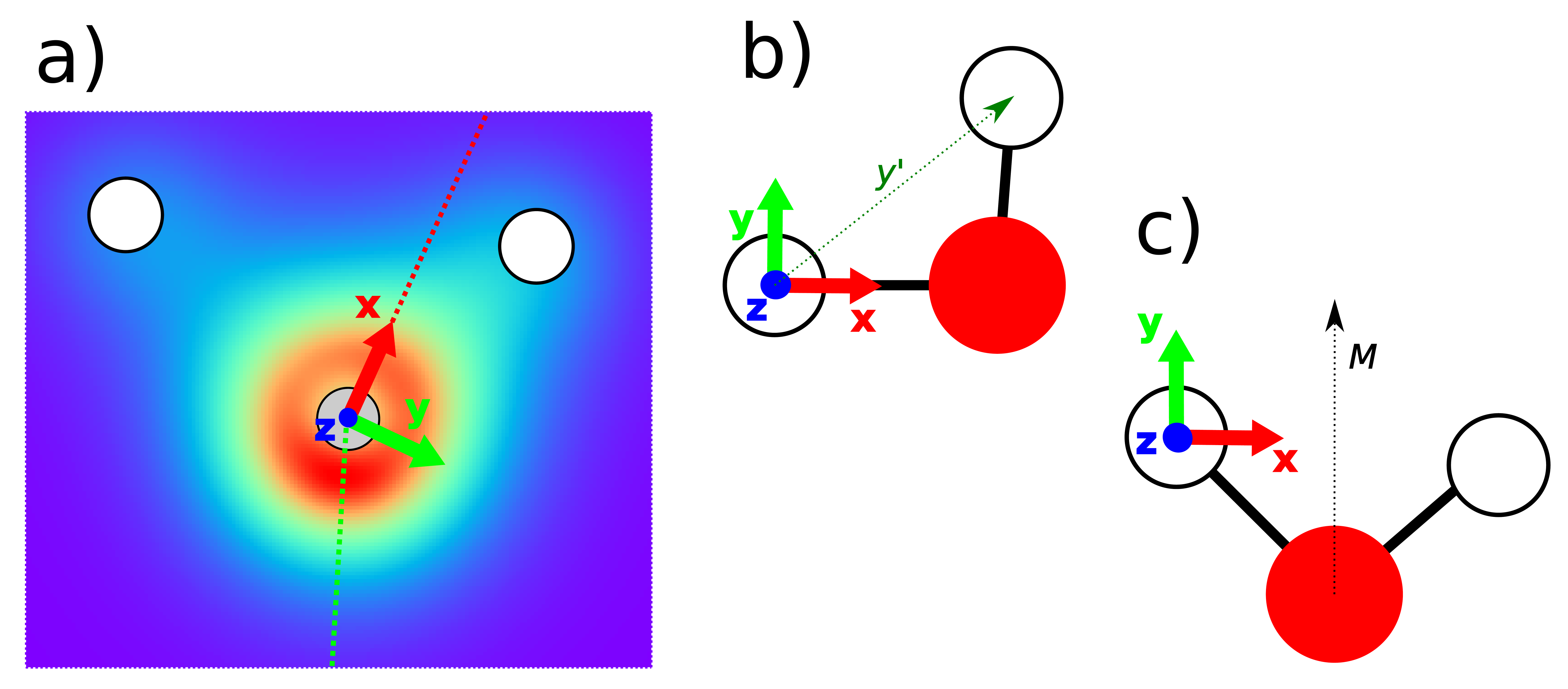}
  \caption{Definition of the a) electronic, b) nearest-neighbor and c) molecular local coordinate system}\label{fig:cs}
\end{figure}
In this work, several definitions of a local coordinate system (LCS) are used.
\paragraph{Electronic}
Given a set of descriptors $\{\tilde c_{nlm}\}$ for a given atom 
the descriptors associated with the p-orbitals $\tilde c_{n1m}$ transform like vectors under $SO(3)$ rotations. 
Starting with $n = 1$ and $l=1$ the corresponding descriptor vector $\tilde c_{11m}$, after being brought into its real form and normalized, is used as the first local axis. Increasing $n$ allows one to find a second vector of $l=1$ descriptors that, using Gram-Schmidt orthogonalization, defines the second local axis.

In the example given in Fig.\ref{fig:cs}a), the first vector $\tilde c_{11m}$ (dashed red line) picks up on the asymmetry of the covalent bonds and is bent towards the closer Hydrogen atom, whereas the second vector $\tilde c_{21m}$ (dashed green line) points towards the lone pair.
Problems may arise if all p-descriptors are collinear, as is the case for highly symmetric systems such as a water monomer with covalent bonds of equal length. In cases like this, one has to revert to using the nearest neighbor rule as described below to determine the second axis. However, extended systems are  rarely this symmetrical and an increase in the number of radial functions is usually enough to resolve issues of linear dependence and find a set of p-descriptors that span a coordinate system. 

\paragraph{Nearest Neighbor}
To determine the LCS around a given atom, one can also use structural information about the atom's immediate environment. The first axis is defined as the direction to the atom's nearest neighbor. The second axis is simply given by the direction to the next nearest neighbor, orthogonalized with respect to axis number one. If all three atoms happen to lie on a line, more distant atoms are taken into consideration until non-colinearity is achieved

\paragraph{Molecular}
For non-reactive models, the molecular axes can be used as a local coordinate system. The axes assignment is hereby arbitrary but following convention we define the y-axis as the bisector (M) of the HOH-angle, the x-axis as the one parallel to the molecular plane and the z-axis orthogonal to it (see Fig.\ref{fig:cs}c)).

\begin{figure}[t]
 \includegraphics[width=.45\textwidth]{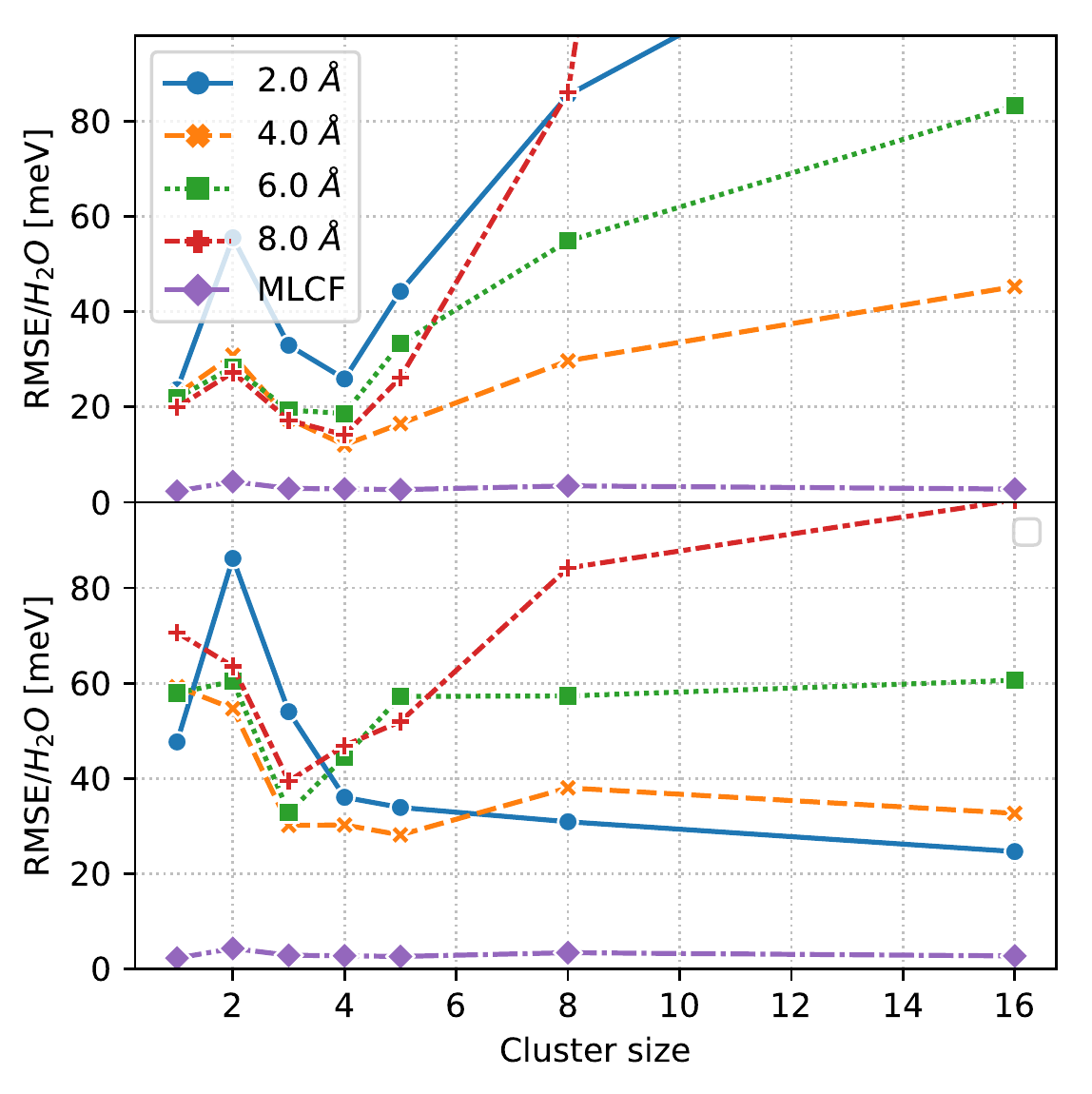}
  \caption{Root mean squared errors (RMSE) in meV per molecule of SchNet (top) and WACSF (bottom) models trained for different cutoff radii (see legend) on water monomers, dimers and trimers and tested on random clusters of increasing size. The errors are compared to those of MLCF (purple line) which can also be found in Tab. \ref{tab:cluster_energies}.}\label{fig:total}
\end{figure}

{
\section{Details on comparison with force fields}

We determined the optimal values for a selected subset of hyperparameters for both SchNet and WACSF by cross-validation (CV), where the validation set was identical to the one used for MLCF.
For SchNet, we fixed the number if Gaussians used to expand the interatomic distance to 25, the dimension of the embedding space (features) as well as the number of interaction layers was determined by CV.
For WACSF the number of fully connected layers in the neural network that acts as an estimator was fixed to three whereas the number of nodes in each layer as well as the number of radial and angular basis functions was determined by CV.
For both SchNet and WACSF a batch size of 100 together with a maximum number of training epochs of 300 was employed. Adam \cite{kingma2014adam} together with a learning rate decay of 0.5 and early stopping was used for optimization.
The optimal set of hyperparameters are shown in Tab. \ref{tab:hyperschnet} and Tab. \ref{tab:hyperwacsf}
for SchNet and WACSF respectively.

Figure \ref{fig:total} shows the performance of non-$\Delta$ models compared to that of MLCF. Even for cluster sizes contained in the training set the force-fields cannot reach competitive accuracy. This is due to the limited size of the training set. It should be noted however, that even though an increase in training set size will most likely lead to a better performance for cluster sizes 1-3, a significant improvement in accuracy for larger clusters is not to be expected as the model will not have enough information to extrapolate to these systems. Better results can be achieved if the force fields are trained to correct the baseline method, as shown in the main text.

}

\begin{table}
\begin{tabular}{c|c|c|c}
\toprule
cutoff & $\Delta$ &  features &  interaction layers \\

\hline
2.0    &        x &       128 &             4 \\
4.0    &        x &       256 &             4 \\
6.0    &        x &       256 &             3 \\
8.0    &        x &       256 &             2 \\
2.0    &          &       256 &             3 \\
4.0    &          &       256 &             3 \\
6.0    &          &       256 &             4 \\
8.0    &          &       256 &             4 \\
\hline
\end{tabular}
\caption{Optimal hyperparameters determined through cross-validation for SchNet. The hyperparameters that were optimized were the the dimension of the embedding space denoted as "features" and the number of 
interaction layers. An "x" in the column named $\Delta$ denotes that the model was trained to correct vdW-cx.}\label{tab:hyperschnet}
\end{table}

\begin{table}
\begin{tabular}{c|c|c|c|c}
\toprule
cutoff & $\Delta$ &  angular &  \# nodes &  radial \\
\hline
2.0    &        x &        8 &      128 &      22 \\
4.0    &        x &        8 &      128 &      22 \\
6.0    &        x &        4 &      128 &      66 \\
8.0    &        x &        8 &      128 &      66 \\
2.0    &          &        8 &      128 &      66 \\
4.0    &          &        8 &      128 &      44 \\
6.0    &          &        4 &      128 &      66 \\
8.0    &          &        6 &      128 &      66 \\
\hline
\end{tabular}
\caption{Optimal hyperparameters determined through cross-validation for WACSF.
The hyperparameters that were optimized were the the number of angular and radial basis functions and the number of nodes in the three layer fully connected neural network. An "x" in the column named $\Delta$ denotes that the model was trained to correct vdW-cx.}\label{tab:hyperwacsf}
\end{table}

\section{Force model}

Given that the mapping from atomic configuration to electron density is obtained by conducting a full self-consistent DFT calculation, the forces cannot be derived from the energy functional, as the derivatives $\partial \Delta E/\partial r_{i,\alpha} = \Delta F_{i,\alpha}$ are unknown.

We therefore have to train a model to directly predict the force corrections $\Delta \vec F$, where again each atomic species is treated with a separate model.
This has the advantage that additional errors that may arise when taking the derivative of the energy can be avoided and models trained directly on forces generally fair better in their prediction than the ones trained on energies.
However, as there is no coupling between energy and force models, the resulting method does not exactly preserve energy and momentum. These effects are due to and thus comparable in magnitude to the fitting error of the force MLCF which in our case is of the order of 0.01 eV/\AA. As an ad-hoc solution for our molecular dynamics simulations we manually set the mean force acting on the system to zero at every time step, making sure that the relative acceleration between Oxygen and Hydrogen atoms remained unaltered. Ongoing research is being conducted into how energy and momentum conservation can be included in the fitting procedure. 

\section{Molecular dynamics}

To test our method's performance in molecular dynamics simulations we trained an MLCF to interpolate between PBE \cite{perdew1996generalized} (baseline) and MB-pol \cite{Babin2013, Babin2014, Medders2014} (reference). The previously employed quadruple zeta doubly polarized was replaced by a faster double zeta polarized (DZP) basis set and the real space mesh cutoff energy was reduced from 300 Ry to 200 Ry. This real space cutoff can be interpreted as the upper limit for the kinetic energy of plane waves that can be represented on this grid without aliasing. 

\begin{figure}[!h]
 \begin{subfigure}[b]{.4\textwidth}
 \caption{}
 \includegraphics[width=\textwidth]{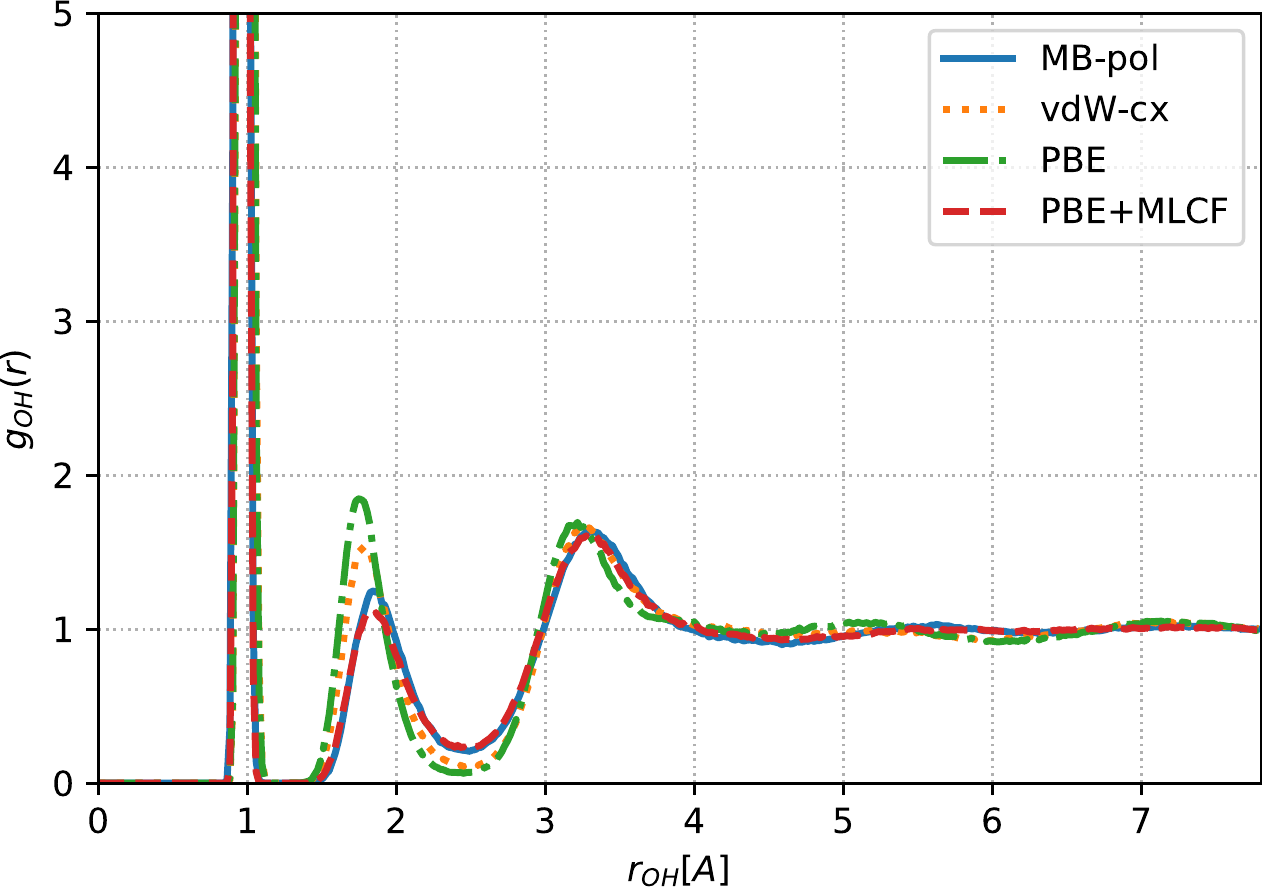}
 \end{subfigure}
 \begin{subfigure}[b]{0.4\textwidth}
 \caption{}
 \includegraphics[width=\textwidth]{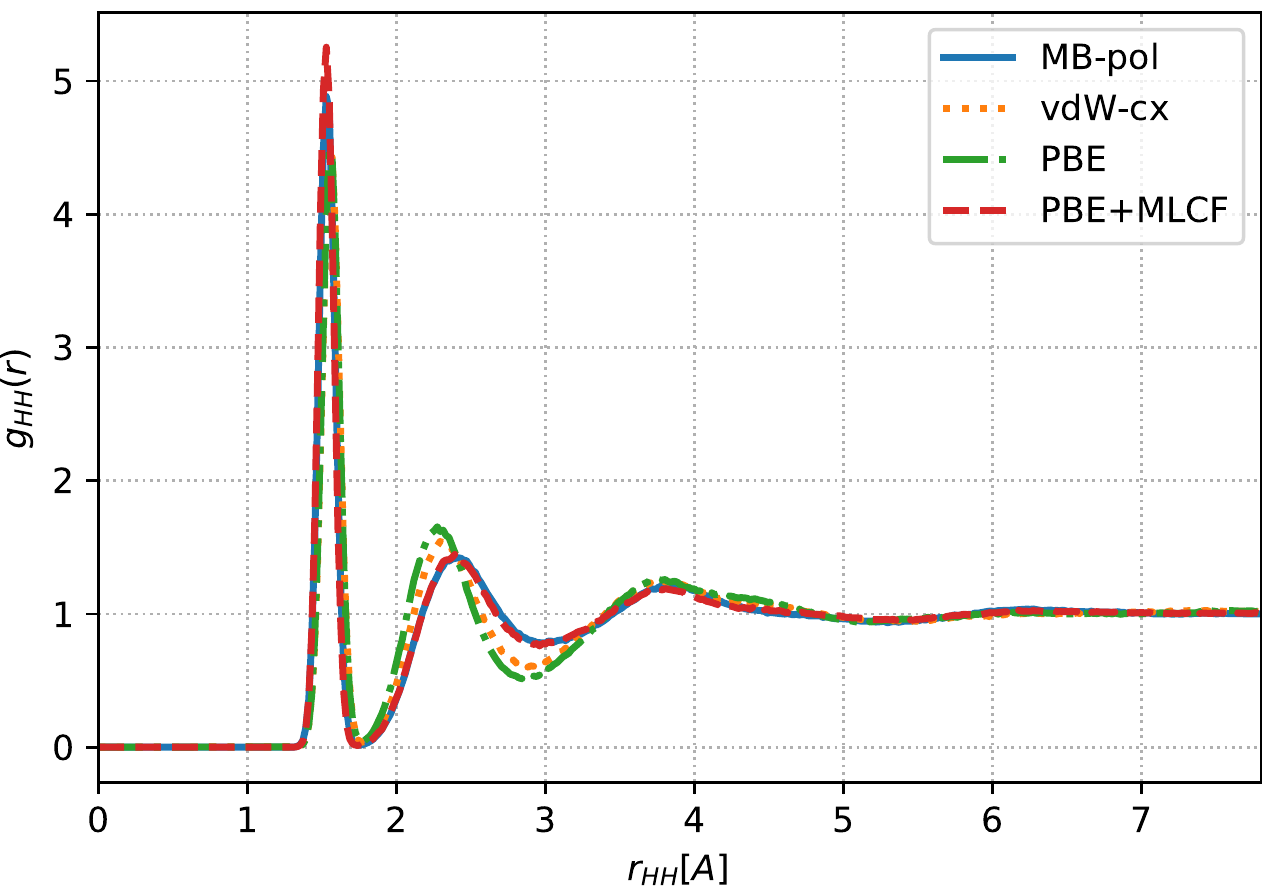}
 \end{subfigure}
 \begin{subfigure}[b]{0.4\textwidth}
 \caption{}
 \includegraphics[width=\textwidth]{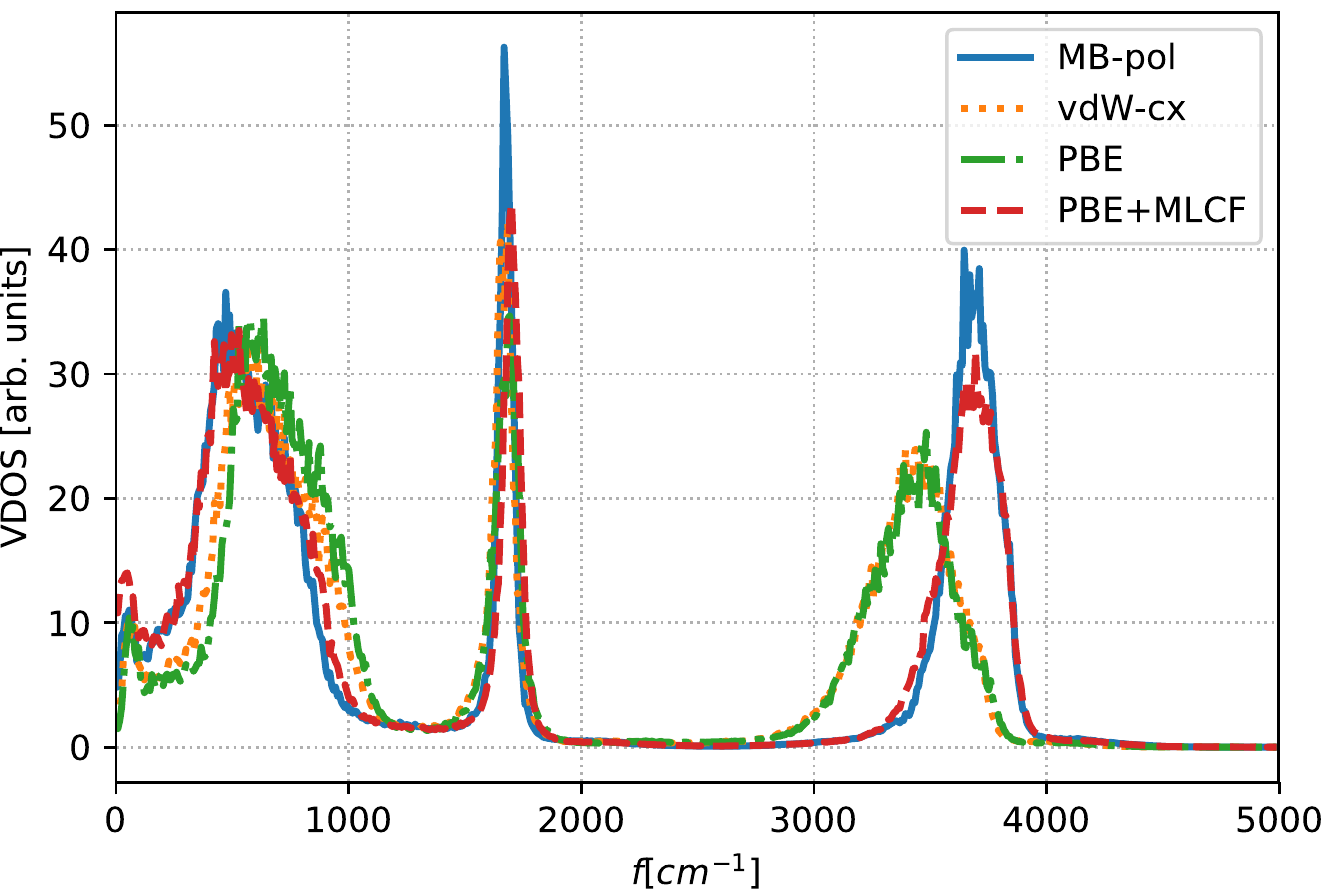}
 \end{subfigure}
  \caption{a) and b) Radial distribution functions and c) vibrational density of states for an MLCF that is trained to correct errors in the XC functional by interpolating between PBE and MB-pol.}\label{fig:rdfmbp}
\end{figure}

\begin{figure}[!h]
 \begin{subfigure}[b]{.4\textwidth}
 \caption{}
 \includegraphics[width=1\textwidth]{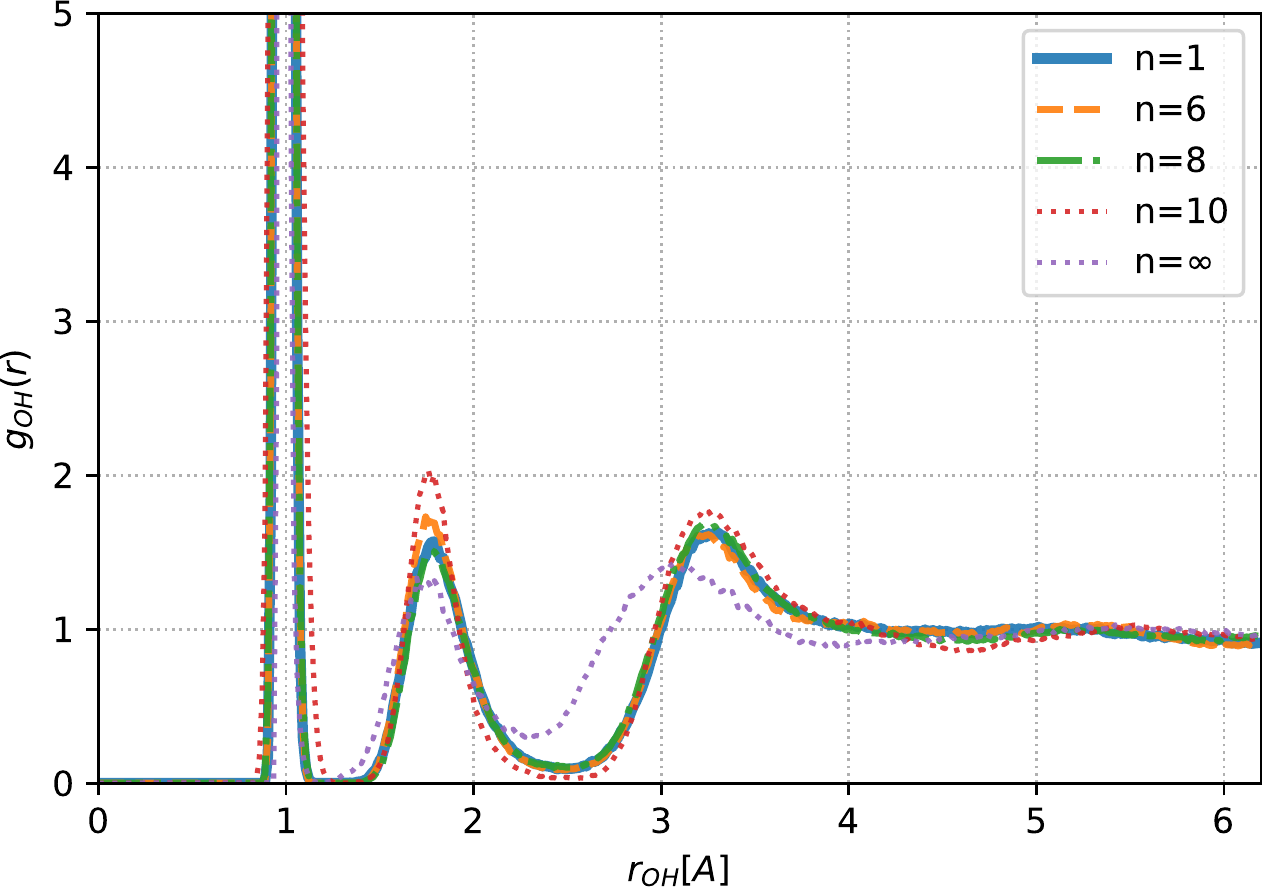}
  \end{subfigure}
  \begin{subfigure}[b]{.4\textwidth}
 \caption{}
 \includegraphics[width=1\textwidth]{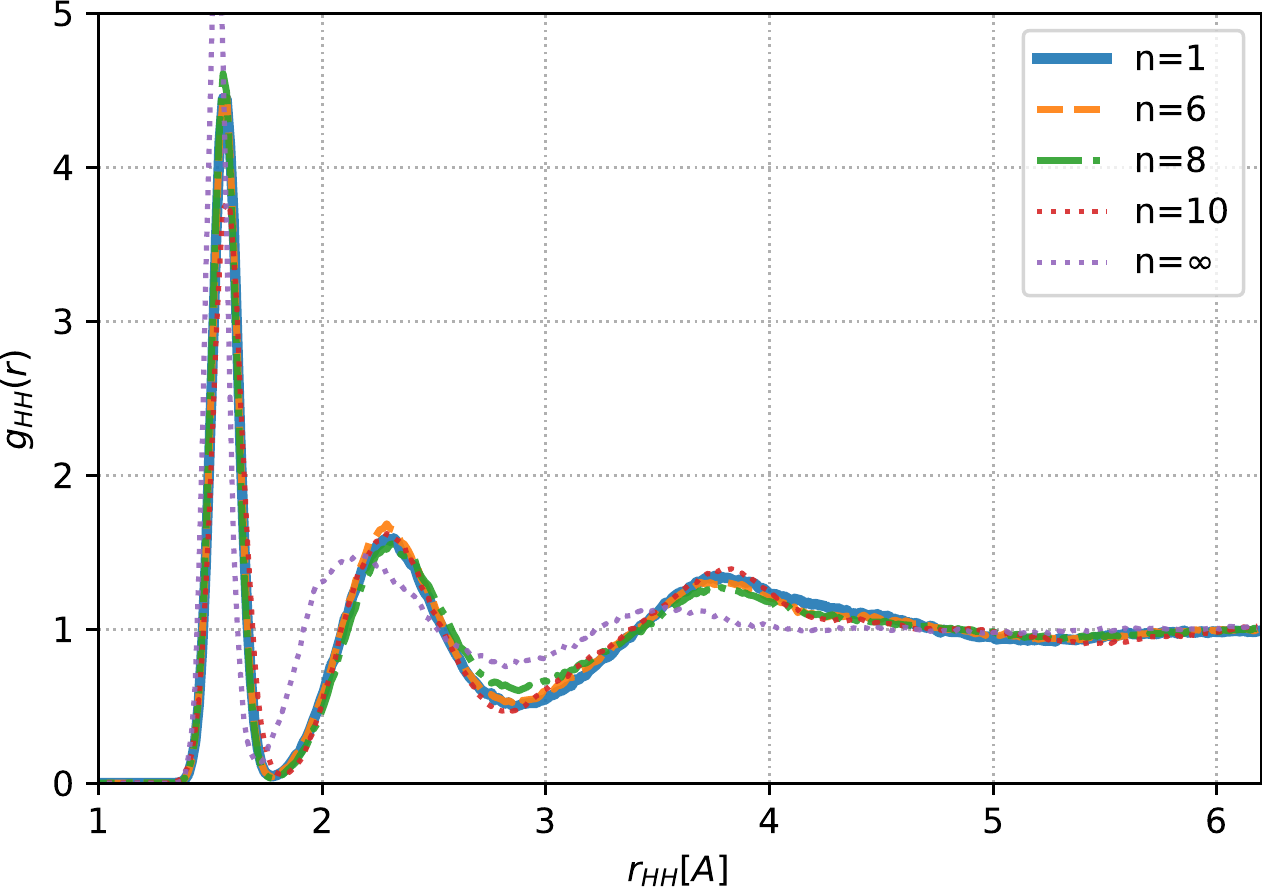}
  \end{subfigure}
  \begin{subfigure}[b]{.4\textwidth}
 \caption{}
 \includegraphics[width=1\textwidth]{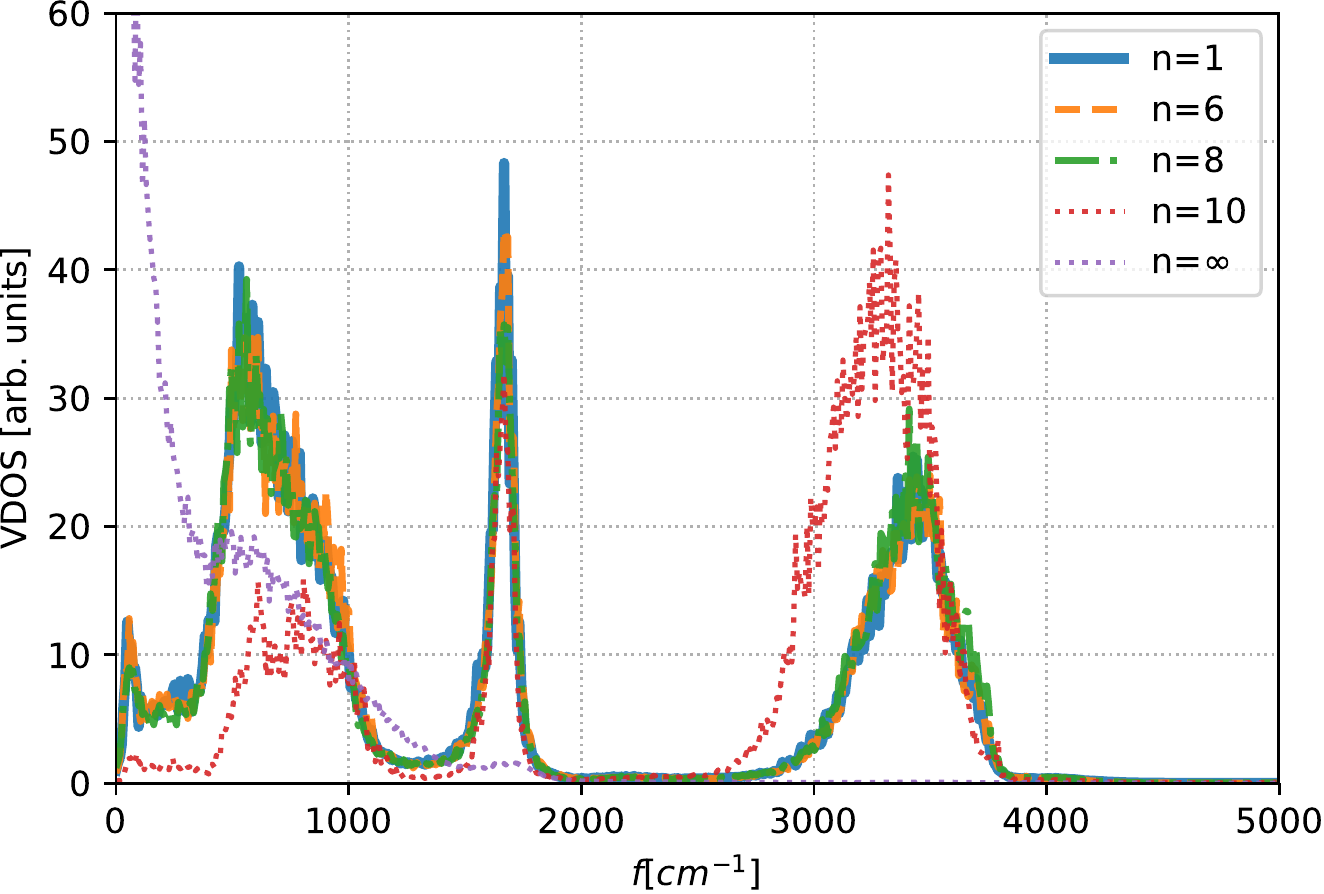}
  \end{subfigure}
  \caption{ a) and b) Radial distribution functions and c) vibrational density of states. MLCF is used to accelerate DFT calculations by correcting for both functional and basis-set related convergence errors, combining MLCF with the time step mixing method proposed by Anglada et al. \cite{anglada2003efficient}. Interpolation between a quick and dirty PBE(SZ) simulation and a slow and more accurate vdW-cx(DZP) simulations. $n$ indicates the number of fast steps needed before correcting with a well converged step.}\label{fig:rdfbh}
\end{figure}
Moving to condensed phase systems, the electron density and thereby the descriptors undergo significant changes, making an MLCF trained on only small clusters inaccurate. The most obvious remedy is to simply train the MLCF on condensed phase data. This, however, has the drawback that  reference calculations become increasingly costly. We avoided this issue by training the model on electron densities of $16(H_2O)$ clusters that were electrostatically embedded in liquid water modeled by TIP4P/2005 \cite{abascal2005general}, opening the possibility of using more expensive, wave function based methods to perform reference calculations in the future.
In detail, this means that a data set was created by sampling a random molecule and its 15 nearest neighbors from snapshots of an MD simulation and treating all remaining water molecules in the snapshot with TIP4P/2005. Doing so, the number of hydrogen bonds that the central molecule forms with its neighbors was used as a stratifying parameter during the sampling. For the descriptor alignment we chose to employ the aforementioned nearest-neighbor rule. To obtain the target energies and forces, the embedding was removed and calculations were performed on clusters only, thereby restricting the correction to short-range effects. Subsequently, only the central molecule was used in fitting the force corrections.

For the MD simulations that used a single zeta (SZ) basis as baseline method we found that a model trained on smaller clusters (monomers, dimers and trimers) exhibited sufficient accuracy. Including larger clusters or using embedding did not lead to any signifant improvement. We accredit this to the 'stiffness' of the SZ basis set which is not capable of describing bonding and polarization effects, making electron densities of small clusters and condensed phases practically indistinguishable. We found, however, that using molecular local coordinate systems (as opposed to nearest neighbor or electronic ones) lead to a slight improvement in model performance.   

In addition to the Oxygen-Oxygen pair correlation function which can be found in the main text, Fig.\ref{fig:rdfmbp} and Fig. \ref{fig:rdfbh} show the OH and HH correlation functions and the vibrational density of states (VDOS) which was obtained by taking the fourier-transform of the velocity autocorrelation function.

\section{Speed-up}

\begin{figure}[t]
 \includegraphics[width=.4\textwidth]{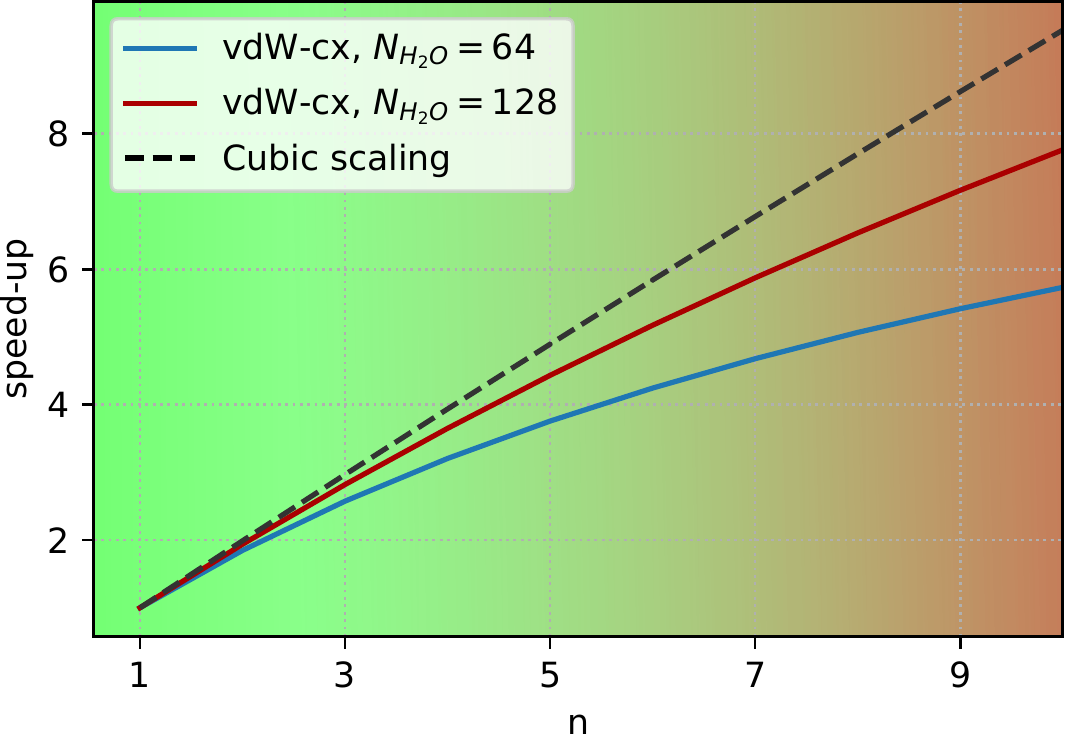}
  \caption{Speed-up obtained by mixing PBE(SZ) with vdW-cx(DZP) for finite size systems compared to the speed-up that would be achieved for an algorithm that scales strictly cubic in the number of oribtals. The color gradient indicates the method's reliability to reproduce the reference method for a given value of $n$.}\label{fig:scaling}
  
\end{figure}

The speed-up that can be achieved with the mixing method by Anglada et al. \cite{anglada2003efficient} depends crucially on both the mixing parameter n and the ratio between the time it takes to propagate the system with the fast method $t_f$ and the time it takes with the slower, accurate method $t_s$.
Alternating (n-1) fast steps with one slow step, the speed-up $\eta$ is given by
\begin{equation}\label{eq:speedup}
    \eta = \frac{n t_{s}}{(n-1)t_{f} + t_{s}}. 
\end{equation}
Fig. \ref{fig:scaling} shows $\eta$ obtained if PBE with SZ basis and vdW-cx \cite{berland2014exchange} with DZP are mixed for which $t_f \approx 35s$ and $t_s \approx 423s$ ($N=64$) and $t_f \approx 89s$ and $t_s \approx 2779s$ ($N=128$) . The values are obtained for calculations on a single  Intel Xeon E5-2683v3 CPU and compared to the theoretical speed-up for an algorithm that is strictly cubic scaling in the number of orbitals. We see that for large enough system sizes $\eta \approx n$.

\bibliography{main}
\end{document}